\documentstyle[12pt]{article}

\begin{document}

\begin{center}

{\large \bf Can Non Gravitational Black Holes Exists?  } \\

\vspace{2.0cm}

M. Novello, V. A. De Lorenci and E. Elbaz$^{\dag}$ \\
\vspace{0.8cm}
{\it Centro Brasileiro de Pesquisas F\'{\i}sicas,} \\
\vspace{0.1cm}
{\it Rua Dr. Xavier Sigaud, 150, Urca} \\
\vspace{0.1cm}
{\it 22290-180 -- Rio de Janeiro, RJ -- Brazil.} \\
\vspace{0.8cm}
$\dag${\it Institut de Physique Nucl\'eaire de Lyon IN2P3-CNRS} \\
\vspace{0.1cm}
{\it Universit\'e Claude Bernard} \\
\vspace{0.1cm}
{\it 43 Bd du 11 Novembre 1918, F-69622 Villeurbanne Cedex, France.}

\vspace{2cm}

\begin{abstract}
We claim that the existence of a mechanism such that photons may be 
trapped in a compact domain is not an exclusive property of gravitational 
forces. We show the case in which a non-linear electrodynamics allows 
such effect. In this latter case we should 
call this region an {\it Electromagnetic Black Hole} (EBH).
\end{abstract}

\end{center}

\vspace{1cm}

Pacs numbers: 04.20.Cv, 04.90.+e 

\newpage

\section{Introduction}

One of the most remarkable consequences of the attractive power of 
gravitational forces is probably the possibility of formation of black holes, 
regions of the spacetime in which photons are confined. 
However, from a fundamental point of view one should be tempted 
to ask: is this a typical and exclusive characteristic of gravity? 
Or, on the other hand, could it be possible that other interactions
present analogous properties, e.g., to display a similar behavior like 
gravitational black holes (GBH), allowing the existence of a hidden part of 
the spacetime structure, unseen from the outside, without making appeal 
to gravity? In this paper we provide an affirmative answer to this question. 
Indeed, we will show that a model can be constructed such that purely 
electromagnetic forces can effectively lead to such a configuration. 
How is this possible?

First of all let us make some comments concerning the structure 
of the metric properties of spacetime in order to introduce our
ideas. Since the advent of general relativity it has been widely accepted 
that the geometry of spacetime is driven uniquely by gravitational forces. 
Although this is a net consequence of the universality of such interaction, 
it is certainly not true that some effects of other interactions cannot be 
described in a similar framework, i.e., such that they can be 
interpreted as being nothing but a modification of the local metric 
properties. 
In order to provide an example -- which will be used as the basis of the whole 
argumentation of the present paper -- let us emphasize that 
it has been known for a long time that the wave propagation of nonlinear 
interactions could well be described in terms of an effective modification 
of the metric qualities of the underlying substractum. To be specific and 
antecipating our result, we shall see that electromagnetic disturbances 
generated in the framework of a nonlinear theory, do not propagate in
an {\it a priori} Minkowskian background structure, but instead, 
propagate in a modified geometry that depends only on the character 
of the nonlinearity of the field. In the case of Born-Infeld theory, for 
instance, it has been shown \cite{Plebansky} that it is the energy-momentum 
density of the non-linear field which is the essential cause by which the 
characteristic surfaces are in general not null cones of the background 
Minkowski geometry but instead null cones of another metric. 
Let us emphasize that the true responsible for the associated 
curvature of this effective geometry has nothing to do with gravitational 
process: it is a pure consequence of the assumed 
non-linearity of the electromagnetic field. We are not 
interested here in pointing out the obvious differences based on the 
distinct property that makes the geometrisation related to gravity 
to be a universal one\footnote{This is due to the validity of the 
equivalence principle.}, in comparison with that produced by 
electromagnetic forces. Notwithstanding its fundamental importance, for our 
purposes here such a distinction is not relevant.

Since all our argumentation in the present paper rests on electrodynamics 
and moreover on its nonlinearity, a few comments on it seems necessary. The 
modification of the metric properties of the underlying geometry through 
which the electromagnetic waves propagate is not an universal phenomenon. 
This means that the effective geometry is nothing but a convenient choice 
of representation of the field propagation in certain 
circumstances\footnote{This 
is not a crucial distinction from general relativity as it can appears in the 
first sight, since, as it has been 
shown (see, for instance refs. \cite{NDL}, \cite{Feynman}), 
Einstein presentation of his gravity theory is nothing but a choice of 
representation. Indeed, all contents of GR can be depicted without appeal to 
the geometrical description. However, due to its universality, 
such (geometric) representation acquired a status of uniqueness. It is 
precisely such universality that makes the most important difference 
between these processes. The equivalence principle guarantees that the 
geometry modification induced by gravity acts in all forms of existing 
matter and energy; on the other hand, only the {\it non-linear photon} 
experiences the modification of the geometry in the non-linear 
electrodynamics.} (see, ref. \cite{NDL} for more details).

We are concerned here with the propagation of disturbances in a 
non-linear electromagnetic theory. This will be taken as a fundamental process. 
However, it seems worth to remark that such method of dealing with an 
equivalent geometry could well be applied to model processes occurring in 
the interior of a medium that contains interactions of charged particles and 
currents like, for instance, in a plasma. The reason for this is simple: 
it is the non-linear nature of the process that is the relevant 
condition for the application of the method of dealing with a modified 
geometry. Thus, it should not be a surprise that some of the consequences 
that we present here could well be tested in a terrestrial laboratory under 
special circumstances. As a major consequence of the equivalent 
geometrical interpretation 
an important property of the theory appears which can be synthesized as:
\begin{itemize}
 \item{{\bf The discontinuities of nonlinear electromagnetic theories 
propagates in an effective non-Minkowskian geometry dependent only on the 
field properties}.}
\end{itemize}
Let us show this and examine some of its consequences.

\section{The General Framework}

The nonlinear electrodynamic theory\footnote{We note that these 
remarks concern any spin-1 theory.} is described by a Lagrangian $L$ 
given uniquely in terms of the invariant 
$F \equiv F_{\mu\nu}\,F^{\mu\nu}$. We set\footnote{We do not consider 
here the invariant constructed with the dual $F_{\mu\nu}^{*}$.}
\begin{equation} 
L = L\,(F). 
\end{equation} 
The corresponding equation of motion is given by 
\begin{equation}
\left\{ L_{F}\, F^{\mu\nu}\right\}_{,\,\nu} = 0
\label{eqmov}
\end{equation}
in which $L_{F}$ represents the functional derivative 
of the Lagrangian ($\delta\,L/ \delta F$)
with respect to invariant $F$; $L_{FF}$ is the second derivative. 

This equation can be written in another, more appealing form, by just 
isolating the linear Maxwell term and taking all remaining non linear 
parts as an additional {\it internal} current to be added to the external one:
\begin{equation}
{F^{\mu\nu}}_{,\,\nu} =   J_{int}^{\mu} + J_{ext}^{\mu} 
\end{equation}
in which the associated {\it internal} current, the self-term is given by 
\begin{equation} 
J_{int}^{\mu} \equiv -\, \frac{L_{FF}}{L_{F}}\,F_{,\nu} \,F^{\mu\nu}.
\end{equation} 
Written under this form it can be thought as nothing but a modelling of 
the response in a self-interacting way of some special plasma medium. 
Indeed, let us consider the quantity $\chi_{\nu}$,
\begin{equation}
\chi_{\nu} \equiv \frac{L_{FF}}{L_{F}}\,F_{,\nu}. 
\end{equation} 
We define a normalized frame $n^{\mu} \equiv \chi^{\mu}/\chi^2$, 
which in the case
$\chi^{\mu}$ is time-like in the Minkowski background, 
could be identified to a real observer that co-moves with $\chi^{\mu}$. 
The extra term of the current assumes then the form 
\begin{equation}
{J^{int}}_{\mu} = \sigma E_{\mu} 
\protect\label{55}
\end{equation}
in which $E_{\mu}$ is the electric part of the field as seen in the 
frame $n^{\mu}$ and $\sigma\, (F)$ may depend on the field variables 
in a complicated way. Under the form of Eq. (\ref{55}) the analogy 
with situations treated within Maxwell electrodynamics in 
material media is transparent. 

From the definition of the energy-momentum tensor we obtain from 
the non-linear Lagrangian:
\begin{equation}
T_{\mu\nu} = - L \gamma_{\mu\nu} - 4 L_{F}\,F_{\mu\alpha}\,{F^{\alpha}}_{\nu}.
\protect\label{56}
\end{equation}

Using the equation of motion and after some manipulation, one obtains the 
expression that contains all information of the balance of forces 
through the exchange of energy of the field and the currents independently of 
the particular form of the Lagrangian. Indeed, we obtain
\begin{equation}
{T^{\mu\nu}}_{,\nu} = - F^{\mu\nu} \,J_{\nu}^{ext}. 
\protect\label{57}
\end{equation}

\subsection{Propagation of the Discontinuities in Non-Linear Electrodynamics}

Let $\Sigma$ be a surface of discontinuity for the electromagnetic field. 
Following Hadamard's \cite{Hadamard} condition let us assume that the field  
is continuous through $\Sigma$ but its first derivative is discontinuous. We 
set
\begin{equation}
[F_{\mu\nu}]_{\Sigma} = 0,
\label{gw1}
\end{equation} 
and   
\begin{equation}
[F_{\mu\nu,\lambda}]_{\Sigma} = f_{\mu\nu} k_{\lambda},
\label{gw2}
\end{equation} 
in which the symbol $[ J ]_{\Sigma}$ represents the discontinuity 
of the function $J$ through the surface $\Sigma$.
Applying these conditions into the equation of motion (\ref{eqmov}) we obtain
\begin{equation}
L_{F}f^{\mu\nu}\, k_{\nu} + 2 L_{FF} \,\xi F^{\mu\nu} k_{\nu} = 0,
\label{gw3}
\end{equation} 
where $\xi$ is defined by 
\begin{equation}
\xi  \equiv  F^{\alpha\beta} \, f_{\alpha\beta}.
\end{equation}
The cyclic identity yields
\begin{equation}
f_{\mu\nu} k_{\lambda} + f_{\nu\lambda} k_{\mu} + 
f_{\lambda\mu} k_{\nu} = 0.
\label{gw4}
\end{equation}
Multiplying this equation by $k^{\lambda}\, F^{\mu\nu}$ yields
\begin{equation}
\xi k_{\nu} \,k_{\mu} \eta^{\mu\nu} + 2 \,F^{\mu\nu} 
f_{\nu\lambda} k^{\lambda} \, k_{\mu} = 0. 
\end{equation}
From the Eq. (\ref{gw3}) it results:
\begin{equation}
f_{\mu\nu} \, k^{\nu} = - \,2\, \frac{L_{FF}}{L_{F}} \, \xi 
F_{\mu\nu} \, k^{\nu},
\end{equation}
and, after some algebraic manipulations the equation of propagation of the
disturbances is obtained:
\begin{equation}
\left\{\gamma^{\mu\nu} + \Lambda^{\mu\nu} \right\} k_{\mu} k_{\nu} = 0.
\label{gww4}
\end{equation}
The new quantity, $\Lambda^{\mu\nu}$, is defined by
\begin{equation}
\Lambda^{\mu\nu} \equiv - 4\, \frac{L_{FF}}{L_{F}} \,
F^{\mu\alpha} \,F_{\alpha}\mbox{}^{\nu}.
\end{equation}

The main lesson we learn from this is that in the non-linear 
electrodynamics the disturbances propagate not in the Minkowskian 
background but in an effective geometry which depends on the energy 
distribution of the field. The net effect of the nonlinearity can 
thus be summarized in the following property.
\begin{itemize} 
\item{The disturbances of nonlinear electrodynamics are null geodesics 
that propagate in the modified effective geometry: 
\begin{equation}
g^{\mu\nu} = \gamma^{\mu\nu}  - 4\, \frac{L_{FF}}{L_{F}} \,
F^{\mu\alpha} \,F_{\alpha}\mbox{}^{\nu}.
\label{geffec}
\end{equation}}
\end{itemize}
In these formulas $\gamma_{\mu\nu}$ is the Minkowski metric written in an 
arbitrary system of coordinates. A simple inspection on this formula shows 
that only in the particular linear case of Maxwell electrodynamics does the 
discontinuity of the electromagnetic field propagate in a Minkowski background.
From equation (\ref{geffec}), we obtain the specific form of the 
components of the metric tensor:
\begin{eqnarray}
g^{00} &=& 1 -  4\, \frac{L_{FF}}{L_{F}} E^2, \\
g^{ij} &=& \gamma^{ij} + 4 \,\frac{L_{FF}}{L_{F}}\,\left(E^{i}\,E^{j} + 
B^{i}\,B^{j} - \gamma^{ij} \,B^{k}\,B_{k}\right), \\
g^{ol} &=& - 4\,\frac{L_{FF}}{L_{F}}\gamma^{li}\epsilon_{ijk}\,E^{j}\,B^{k},
\end{eqnarray}
in which we have set $E^2 = -\, E_{\alpha}\,E^{\alpha}$.

For future references we note that in the case we will be concerned here, 
the ratio $\frac{L_{FF}}{L_{F}}$ is a constant. Thus, the $g_{00}$ component 
of the effective geometry depends only on the electric part of the field, 
while the off-diagonal terms $g^{0l}$ are directly related to the Poynting
vector. We can thus exhibit the conditions in which the effective metric can 
be written in a Gaussian system of coordinates: the vanishing of the 
Poynting vector and the constancy of the norm of the electric field. 

A direct inspection on the above formulas of the associated 
geometry allows us to envisage the
possibility of generating a null surface exclusively in 
terms of electromagnetic processes. This situation happens to occur 
when the properties of the non linearity is such that it induces the 
equality
\begin{equation}
E^2 = L_{F}/4L_{FF}.
\end{equation}
Before going into a specific model, let us make here a comment.
Linear photons propagate in a Minkowskian underlying background.
Non linear photons propagate in an effective geometry given by
Eq. (\ref{geffec}). 
Note, however, that this situation is not competitive to gravity
processes. The reason for this is easy to understand: the above modified
geometry (in case of non-linear electrodynamics) is not a universal one. 
Indeed, other kinds of particles and radiations behave as if the 
background metric is that dealt with in special relativity: 
the charged particles, e. g., electrons follow time-like paths 
with respect to Minkowski metric.

In the appendix we present a simple example of the application of
such procedure in the case of electrodynamics in a dielectric medium.

\section{The Model}

In order to show a specific situation that represents a configuration 
of an effective geometry for electromagnetic forces, let us concentrate here 
in a simple model. We set
\begin{equation}
L  = -\frac{F}{4}\left(1-\frac{F}{b}\right)^{-1},
\label{Lmodel}
\end{equation}
where the constant $b$ has dimensionality of density of energy (we use 
units in which $c = 1$). We should note that this theory is a
very close approximation of Maxwell electrodynamics in the case 
the constant $b$ is large.

For our purposes, it is convenient to seek for a spherically symmetric and 
static solution of this theory. A direct computation 
shows that in the spherical coordinate system $(t, r, \theta, \phi)$ a 
particular solution can be found uniquely in terms of a radial electric 
component given by:
\begin{eqnarray}
F_{01} &=& f(r),
\end{eqnarray}
where the function $f(r)$ obeys the relation:
\begin{equation}
\frac{f(r)}{\left[b+2f(r)^2\right]^2} = \frac{C_{0}}{r^2}
\label{relationf}
\end{equation}
and the parameter $C_{0}$ is related to the electric charge located at the
origin.

From the previous section, we conclude that the effect of the 
nonlinearity on the 
propagation of the discontinuities is to induce the electromagnetic
waves to follow a null cone of the modified geometry given by
\begin{equation}
g^{\mu\nu} = \gamma^{\mu\nu} - \frac{8}{b-F}\, F^{\mu\beta} 
F_{\beta}\mbox{}^{\nu},
\end{equation}
in which the Minkowski metric $\gamma_{\mu\nu}$ has the form
\begin{equation}
ds^2 = dt^2 - dr^2 - r^2\,d\,\theta^2 - r^2\, \sin^2\theta \,d\phi^2.
\end{equation}
It then follows that the non-null components of the metric tensor of 
the effective geometry, as seen by the photon disturbances, are:
\begin{eqnarray}
g^{00} &=& \frac{b-6f(r)^2}{b+2f(r)^2},\\
g^{11} &=& -\, g^{00},\\
g^{22} &=& -\, \frac{1}{r^2},\\
g^{33} &=& \frac{1}{r^2 \,\sin^2\theta}.
\end{eqnarray}

Just for completeness, let us exhibit the non-vanishing components 
of the energy-momentum tensor of the field:
\begin{eqnarray}
T^{0}\mbox{}_{0} &=& \frac{bf(r)^2}{2}\left\{\frac{b-2f(r)^2}{[b+2f(r)^2]^2}\right\}, \\
T^{1}\mbox{}_{1} &=& T^{0}\mbox{}_{0},\\
T^{2}\mbox{}_{2} &=& -\frac{bf(r)^2}{2b+4f(r)^2}, \\
T^{3}\mbox{}_{3} &=&  T^{2}\mbox{}_{2}.
\end{eqnarray}
In linear Maxwell theory the energy-momentum tensor is traceless. This 
property is no longer true in the non-linear case. A direct inspection on this 
expression shows that the field energy is well behaved throughout 
all space except at the origin.

\subsection{Electromagnetic Black Holes (EBH)}

Electromagnetic disturbances in a non-linear theory follow null cones 
(geodesics) in an effective geometry. The best way to analyze the 
properties of their paths is then to examine the equations of motion of the 
geodesics in the effective metric $g_{\mu\nu}$ given by
\begin{equation}
ds^2 = \left[\frac{b+2f(r)^2}{b-6f(r)^2}\right]dt^2 - 
\left[\frac{b+2f(r)^2}{b-6f(r)^2}\right]dr^2 - r^2 \,d\,\theta^2 - r^2\, 
\sin^2\theta \,d\phi^2.
\label{effmetric}
\end{equation}
Solving the variational problem 
\begin{equation}
\delta\int ds = 0,
\end{equation}
the solutions of the Euler-Lagrange equations that follow are: 
\begin{eqnarray}
\left[\frac{b+2f(r)^2}{b-6f(r)^2}\right]\dot{t} &=& l_0,
\label{l0}
\\
r^2\dot{\phi} &=& h_0,
\label{h0}
\end{eqnarray}
where $l_0$ and $h_0$ are constants of motion related to the photon energy. 
The system was reduced to a planar orbit by the choice of the initial 
conditions:
\begin{eqnarray}
\dot{\theta} &=& 0,
\nonumber\\
\theta &=& \frac{\pi}{2}.
\label{cond}
\end{eqnarray}
The null property of the geodesics allows to obtain in a direct way 
the equation of the radial component. Using Eqs. (\ref{effmetric}) -
(\ref{cond}), results:
\begin{equation} 
\dot{r}^2 + \frac{b-6f(r)^2}{\left[b+2f(r)^2\right]^2}
\left\{\frac{h_{0}^2f(r)}{C_0} - l_{0}^2\left[b-6f(r)^2\right]\right\} = 0.
\label{radialeq}
\end{equation}
We can re-write this equation in the more convenient form:
\begin{equation} 
\dot{r}^2 + V_{eff} = l_{0}^2
\end{equation}
in which the potential $V_{eff}$ is defined by:
\begin{equation} 
V_{eff} = \frac{b-6f(r)^2}{\left[b+2f(r)^2\right]^2}
\left\{\frac{h_{0}^2f(r)}{C_0} - l_{0}^2\left[b-6f(r)^2\right]\right\}
+ l_{0}^2.
\label{potential}
\end{equation}
Thus, the motion of the photon in such non-linear theory can be 
described as a particle dotted with energy $l_{0}^2$ immersed in a 
central field of forces characterized by the potential $V_{eff}$.

\section{Conclusion}

The important fact to be noticed here is related to the behavior of
the metric (\ref{effmetric}) in the critical point $f(r_{c})^2 = b/6$.
Indeed coordinates $t$ and $r$ interchange their corresponding roles 
when crossing the $r=r_{c}$ null surface\footnote{See appendix B for more
details.}. Note however that the potential $V_{eff}$ is well 
behaved at $r_{c}$. 
 
We can thus conclude from this remark and from what we have learned in this 
paper, that configurations of structures containing hidden regions, like black
holes, are not restricted to gravitational forces. Non linear electromagnetic 
interaction can also produce similar objects. 
In this paper we presented a simple purely electromagnetic model in which we 
have neglected the gravitational effects. 

We should like to point out that in order to compare the above EBH 
(Electromagnetic Black Hole) to
standard GBH (Gravitational Black Hole) one must couple this non
linear electromagnetic theory to gravity. We postpone this 
analysis for a forthcoming paper.

\section{Acknowledgements}

We would like to thank the participants of the ``Pequeno Semin\'ario''
of the cosmology group of Lafex/CBPF for some comments on this paper. 
This work was supported by ``Conselho Nacional de Desenvolvimento Cient\'{\i}fico
e Tecnol\'ogico (CNPq)'' of Brazil.

\section*{Appendix}

\subsection*{A -- Dielectric Constant and the Effective Geometry}
\setcounter{equation}{0}
\def\theequation{A.\arabic{equation}}

In order to show the non-familiar reader the treatment that involves  
dealing with the propagation of the non-linear theory as a modification 
of the background geometry, we will present here the simplest possible case 
of the standard Maxwell theory in a dielectric medium. We will show how 
it is possible to present the wave propagation of linear electrodynamics in 
a medium in terms of a modified geometry of the spacetime.

In this section we take the Maxwell theory in a medium such that the 
electromagnetic field is represented by two anti-symmetric tensors 
$F_{\mu\nu}$ and $P_{\mu\nu}$ given in terms of the electric and 
magnetic vectors, as seen by an arbitrary observer 
endowed with a four-velocity $v^{\mu}$, by the standard expressions:
\begin{equation}
F_{\mu\nu} = E_{\mu}\, v_{\nu} - E_{\nu}\, v_{\mu} + 
\eta_{\mu\nu}\mbox{}^{\rho\sigma} \, v_{\rho} \, H_{\sigma}
\end{equation}
and
\begin{equation}
P_{\mu\nu} = D_{\mu}\, v_{\nu} - D_{\nu}\, v_{\mu} + 
\eta_{\mu\nu}\mbox{}^{\rho\sigma} \, v_{\rho} \, B_{\sigma}.
\end{equation}
The Maxwell equations are:
\begin{equation}
\partial^{\nu} \,F^{*}_{\mu\nu} = 0,
\end{equation}
\begin{equation}
\partial^{\nu} \,P_{\mu\nu} = 0.
\end{equation}
Following Hadamard, we consider the discontinuities on the fields 
as given by
\begin{eqnarray}
\left[\partial_{\lambda}\, E_{\mu}\right]_{\Sigma} &=& k_{\lambda}\, 
e_{\mu}
\nonumber\\
\left[\partial_{\lambda}\, D_{\mu}\right]_{\Sigma} &=& k_{\lambda}\, 
d_{\mu}
\nonumber\\
\left[\partial_{\lambda}\, H_{\mu}\right]_{\Sigma} &=& k_{\lambda}\, 
h_{\mu}
\nonumber\\
\left[\partial_{\lambda}\, B_{\mu}\right]_{\Sigma} &=& k_{\lambda}\, 
b_{\mu}.
\end{eqnarray} 
Using the constitutive relations\footnote{We deal here with the
simplest case of linear isotropic relations, just for didactic reasons.}
\begin{eqnarray}
d_{\mu} &=&  \epsilon \, e_{\mu} 
\\
b_{\mu} &=&   \frac{h_{\mu}}{\mu} 
\end{eqnarray}
one obtains after a straightforward calculation
\begin{equation}
k_{\mu}\, k_{\nu} \left[ \eta^{\mu\nu} + (\epsilon\, \mu - 1) v^{\mu}
\,v^{\nu} \right] = 0.
\end{equation}

This shows that even the simple case of the evolution of the 
wave front in standard Maxwell equation in a medium can be
interpreted in terms of an effective geometry $g^{\mu\nu}$ that
depends not only on the medium properties $\epsilon$ and $\mu$, but 
also on the observer\rq s velocity, given by: 
\begin{equation} 
g^{\mu\nu} \equiv  \eta^{\mu\nu} + (\epsilon\, \mu - 1) v^{\mu}
\,v^{\nu}.
\end{equation}
This ends our proof.

\subsection*{B -- Null Surface}
\setcounter{equation}{0}
\def\theequation{B.\arabic{equation}}

Let us consider the surface $\psi=r=const$ in the 
case of the solution examined in the previous section. 
We are interested here in
the analysis of the characteristics of the  equation of motion of the
non linear electromagnetic  field in the neighborhood of the critical 
radius defined by relation 
\begin{equation}
f(r=r_{c}) = \sqrt{\frac{b}{6}}
\end{equation}
that result in the value for this radial coordinate
\begin{equation}
r_{c} = \frac{4b}{3}\sqrt{C_{0}\sqrt{\frac{6}{b}}}.
\end{equation} 
Using the metric $g^{\mu\nu}$, we have
\begin{equation}
\psi_{\mu}\psi_{\nu}g^{\mu\nu}=\psi_{1}\psi_{1}g^{11}=
-(\psi_{1})^2\left[\frac{b-6f(r)^2}{b+2f(r)^2}\right]
\end{equation}
where we have set $\psi_{\mu} = \partial_{\mu}\psi$. 
At the value $r = r_{c}$, this relation vanishes showing that
the surface $\psi$ is a null surface at the critical radius, for 
the non-linear photon.

\subsection*{C -- Born-Infeld Model}
\setcounter{equation}{0}
\def\theequation{C.\arabic{equation}}

The result that we have shown here, concerning the existence of EBH depends 
not only on the nonlinearity but also on the specific form of theory.
The most popular non-linear model, the Born-Infeld \cite{Born} 
theory, does not admit such structure (EBH) in its corresponding 
static configuration. Indeed, the simplest way to show this is by a 
direct inspection on the properties of the
corresponding static solution for this case. The Born-Infeld Lagrangian is 
\begin{equation}
L = -\frac{b^2}{4} \left\{ \sqrt{1 + \frac{2\,F}{b^2}} - 1 \right\}. 
\end{equation}
From the previous formuli, the $g^{00}$ component of the effective 
geometry takes the form
\begin{equation} 
g^{00} = 1 + \frac{4E^2}{b^2 - 4\,E^2} 
\end{equation} 
that results in the inverse --- since in this solution we have the metric
diagonal --- given by
\begin{equation} 
g_{00} = \frac{b^2 - 4\,E^2}{b^2}
\end{equation}
which vanishes for the following value of the electric field:
\begin{equation}
E = \frac{b}{2}.
\end{equation}
Therefore, this value of the electric field is just the upper limit
of validity of this function, as it can be noticed by a direct analysis
of the above Lagrangian.

\end{document}